\documentclass[aps,pre,preprint,groupedaddress]{revtex4}

\usepackage[dvips]{graphicx}
\usepackage{mathrsfs}
\usepackage{amsmath,amssymb}
\usepackage{hhline}
\usepackage{dcolumn}
\usepackage{bm}

\usepackage[dvips]{color}

\begin{document}

\title{Hierarchical scale-free network is fragile against random failure}

\author{Takehisa Hasegawa}
 \email{hasegawa@m.tohoku.ac.jp}
\affiliation{%
Graduate School of Information Science, 
Tohoku University, 
6-3-09, Aramaki-Aza-Aoba, Sendai, Miyagi, 980-8579, Japan
}%
\author{Koji Nemoto}
\email{nemoto@statphys.sci.hokudai.ac.jp}
\affiliation{%
Department of Physics, Hokkaido University,
Kita 10 Nisi 8, Kita-ku, Sapporo, Hokkaido, 060-0810, Japan
}%

\begin{abstract}
We investigate site percolation in a hierarchical scale-free network known as
the Dorogovtsev-Goltsev-Mendes network.
We use the generating function method to show that the percolation threshold is 1, 
i.e., the system is not in the percolating phase when the occupation probability is less than 1.
The present result is contrasted to bond percolation in the same network of which the percolation threshold is zero. 
We also show that the percolation threshold of intentional attacks is 1. 
Our results suggest that this hierarchical scale-free network is very fragile against both random failure and intentional attacks. 
Such a structural defect is common in many hierarchical network models.
\end{abstract}

\pacs{89.75.Hc,87.23.Ge,05.70.Fh,64.60.aq}

\maketitle



\section{Introduction}

The prominent resilience of real networks to random failure and intentional attacks is one of the important issues in network science
\cite{albert2002statistical,newman2003structure,dorogovtsev2008critical,barrat2008dynamical}.
Many real networks are scale free, i.e., the degree distribution $p(k)$ is a power law denoted by $p(k) \propto k^{-\gamma}$ with $2 \le \gamma \lesssim 3$. 
Albert {\it et al.} examined the robustness of networks against two types of attacks: 
random failure in which nodes are sequentially removed with equal probability 
and intentional attack, which preferentially removes nodes of large degrees \cite{albert2000error}. 
They showed that scale-free networks with small $\gamma$ are highly robust against random failure, 
i.e., the network remains intact until almost all nodes have been removed.
On the other hand, such networks are very fragile to intentional attacks because removal of a small fraction of hubs destroys the network.

Random failures and intentional attacks in networks can be interpreted as percolation problems and have been well studied 
(see Refs.~\cite{newman2003structure,dorogovtsev2008critical} and references therein). 
It is well known that the site (bond) percolation model with the probability $p$ that 
each site (bond) is occupied (open) has a percolation threshold $p_c$ 
above which the largest connected component is $O(N)$, $N$ being the number of nodes.
Such a network is said to be {\it robust} ({\it fragile}) against failure if $p_c$ $(1-p_c)$ is very small
when a random failure is regarded as a node vacancy in site percolation and as a closed link in bond percolation. 
The local tree approximation for uncorrelated networks \cite{molloy1995critical}, 
which is a standard theory in network science \cite{dorogovtsev2008critical}, 
confirms that scale-free networks with heavy-tailed degree distributions are robust against random failures \cite{callaway2000robustness,cohen2000resilience}, 
i.e., $p_c$ is zero for both bond and site percolations when $\gamma \le 3$.
This approximation can be applied to the case of intentional attacks 
to show that uncorrelated scale-free networks with small $\gamma$ are fragile against such attacks \cite{callaway2000robustness,cohen2001breakdown}. 
This theory can be extended so as to treat clustered networks and correlated networks \cite{goltsev2008percolation,tanizawa2012robustness,newman2003properties,
gleeson2009bond,gleeson2009analytical,gleeson2010how}.

\begin{figure}[b]
\begin{center}
\includegraphics[width=60mm]{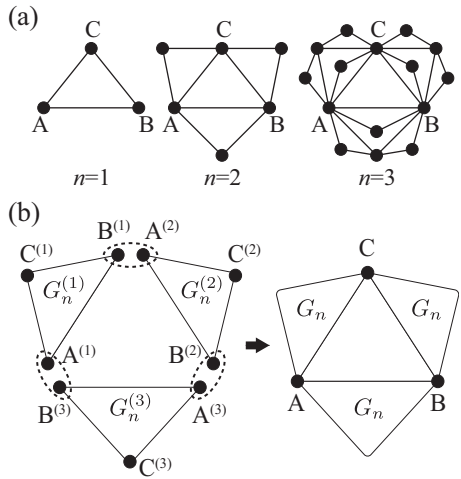}
\end{center}
\caption{
(a) Realization of the DGM network with generation $n=1$, $2$, and $3$. 
(b) Construction of $G_{n+1}$ from $G_n$. Three copies of $G_n$ ($G_n^{(k)}, k=1,2,3$)
are connected by identifying 
A$^{(1)}$ and B$^{(3)}$ to be the new A, A$^{(3)}$ and B$^{(2)}$ to be the new B, and A$^{(2)}$ and B$^{(1)}$ to be the new C, 
respectively.
}
\label{fig-construction}
\end{figure}

Apart from the local tree approximation, 
bond percolation in growing and hierarchical networks has been studied extensively (see below), 
whereas, almost no analytical studies have focused on site percolation.
To date, little attention has been paid to
the difference between bond percolation and site percolation in complex networks,
possibly because the critical properties of the two percolation models are not qualitatively different within
the local tree approximation \cite{callaway2000robustness,cohen2000resilience}. 
However, what we will demonstrate here is that the opposite can occur. 
We examine site percolation in a hierarchical scale-free network
known as the Dorogovtsev-Goltsev-Mendes (DGM) network \cite{dorogovtsev2002pseudofractal} or the (1,2) flower \cite{rozenfeld2007fractal,rozenfeld2007percolation}.
Dorogovtsev \cite{dorogovtsev2003renormalization} calculated bond percolation in this network by renormalization group
and showed that the percolation threshold $p_c$ is zero. 
On the other hand, we use generating functions to show that 
the percolation threshold $p_c$ of site percolation is not zero but one. 
That is, this network is fragile even against random failure.
Our analytical result is supported perfectly by Monte Carlo simulation.

Before discussing the main topic, we emphasize that 
percolation characteristics on some graphs are delineated by two nontrivial transition points, 
namely, $p_{c1}$ and $p_{c2}$ \cite{benjamini1996percolation,lyons2000phase,schonmann2001multiplicity}. 
According to the value of $p$, the system shows one of the three phases: 
(i) nonpercolating phase ($0 \le  p \le p_{c1}$) in which all clusters are of finite size, 
(ii) critical phase ($p_{c1} < p < p_{c2}$) (called the patchy phase in Ref.~\cite{boettcher2009patchy}) 
in which infinitely many infinite clusters exist, 
and (iii) percolating phase ($p_{c2} \le p \le  1$) in which the system has a unique infinite cluster. 
Here an infinite cluster is a cluster whose size is on the order $O(N^\alpha) (0 < \alpha \le 1)$.
Note that $p_{c2}$ is equal to the percolation threshold $p_c$, $p_c=p_{c2}$. 
By using the order parameter $m \equiv \lim_{N \to \infty} s_{\rm max}(N; p)/N$
and the fractal exponent $\psi \equiv \lim_{N \to \infty} \log_N s_{\rm max}(N)$ \cite{nogawa2009monte,hasegawa2010profile}, 
we represent these phases as (i) $m=0$ and $\psi=0$, (ii) $m=0$ and $0<\psi<1$, and (iii) $m>0$ and $\psi=1$, respectively.
Here $s_{\rm max}(N; p)$ is the mean size of the largest cluster in a graph of size $N$ at a given value of $p$.
As known, $p_{c1}=p_{c2}$ on Euclidean lattices, 
whereas, $p_{c1}<p_{c2}$ on transitive nonamenable graphs \cite{lyons2000phase,schonmann2001multiplicity}.
Also, in complex networks, 
some growing network models \cite{callaway2001randomly,dorogovtsev2001anomalous,zalanyi2003properties,hasegawa2010critical,hasegawa2010profile} 
and hierarchical network models \cite{boettcher2009patchy,berker2009critical,boettcher2012ordinary,hasegawa2010generating,hasegawa2012absence} 
yield $0=p_{c1}<p_{c2}$ for bond percolation,
whereas, $p_{c1}=p_{c2}$ for some static network models such as uncorrelated networks \cite{newman2003structure,dorogovtsev2008critical}.
In this paper, we analytically show that $p_{c1}=0$ and $p_{c2}=1$ for site percolation, 
whereas, $p_{c1} = p_{c2} = 0$ for bond percolation \cite{dorogovtsev2003renormalization}.
We also demonstrate that $p_{c1}=p_{c2}=1$ for intentional attacks on the DGM network. 


\section{Model}

The DGM network was proposed as a deterministic growing network \cite{dorogovtsev2002pseudofractal}. 
Let us denote the DGM network with generation (= time) $n$ by $G_n$. 
The model starts from a triangle at $n=1$. 
At each time step $n$, every link in $G_n$ adds a new node, which links to both end nodes of the link, to create $G_{n+1}$. 
The realizations of the first three generations, $G_1$, $G_2$, and $G_3$, are shown in Fig.~\ref{fig-construction}(a). 
This model also is regarded as a recursively constructed hierarchical network: 
$G_1$ consists of a triangle of nodes A, B, and C. We refer to these nodes as {\it roots}. 
Then, $G_{n+1}$ is constructed from three copies of $G_{n}$ that are joined at the roots as shown in Fig.~\ref{fig-construction}(b).

The structural properties of this network have been described in Refs.~\cite{dorogovtsev2002pseudofractal,rozenfeld2007fractal}.
The number of links in $G_n$ is $3^n$ and the number of nodes is $(3^n+3)/2$.
The number of nodes $N_n(\ell)$ of degree $k=2^\ell$ $(\ell=1,2,\ldots,n)$ is $N_n(\ell)=3^{n-\ell}$ for $\ell<n$ and $N_n(n)=3$. 
Thus, the degree distribution $p(k)$ is a power law, $p(k) \propto k^{-\gamma}$ with $\gamma = 1 + \ln 3 / \ln 2$.
Furthermore, the DGM network is a small-world network because the diameter of $G_n$ is $n$ and the clustering coefficient is $4/5$ in the limit $n \to \infty$.

\begin{figure}
 \begin{center}
  \includegraphics[width=50mm]{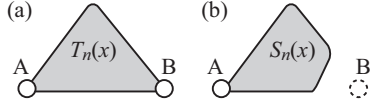}
 \end{center}
 \caption{
Schematic of (a) $T_n(x)$ and (b) $S_n(x)$.
}
\label{fig-diagramA}
\end{figure}

\begin{figure}
 \begin{center}
  \includegraphics[width=75mm]{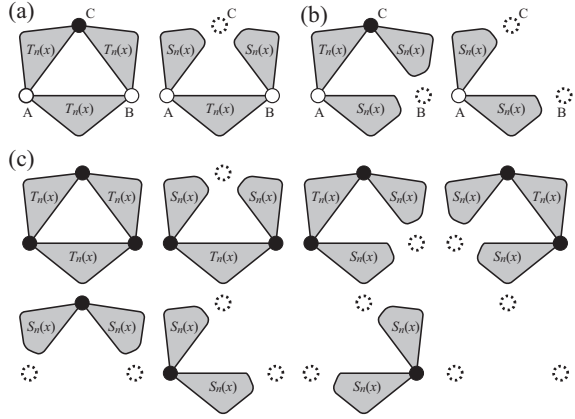}
 \end{center}
 \caption{
Possible contributions to (a) $T_{n+1}(x)$, (b) $S_{n+1}(x)$, and (c) $F_{n+1}(x)$. 
Each solid (dashed) circle represents the root node being occupied (unoccupied). 
The black circles represent occupied root nodes which should be taken into account by multiplying $x$. 
For example, the first diagram of (a) represents $p x T_n^3(x)$, the second diagram of (b) represents $q S_n^2(x)$, 
and the second one in the second line of (c) represents $p q^2 x S_n^2 (x)$.
}
\label{fig-diagramB}
\end{figure}



\section{Generating function}

Let us consider site percolation in $G_n$ with an occupation probability $p$. 
We calculate the mean size $s_{\rm root}(N_n;p)$ of the root cluster, 
which includes at least one of the roots A, B, and C, 
rather than the mean largest cluster size $s_{\rm max}(N_n;p)$. 
Here, A, B, and C have the largest degree $k=2^n$ in $G_n$ and can, therefore, be regarded as {\it hubs}.
Also, the root cluster is always unique because these roots are connected directly in the cluster.
Since this root cluster is expected to become the largest cluster, we assume that 
$s_{\rm max}(N_n;p)$ is well approximated by $s_{\rm root}(N_n;p)$.
This assumption is verified numerically below.

To evaluate $s_{\rm root}(N_n;p)$, we consider the following two quantities in $G_n$: 
The probability that the size of the root cluster is $k$ provided that both A and B are occupied 
(denoted as $t_k^{(n)}(p)$, we call such clusters doubly-occupied),
and the probability that the size of the root cluster is $k$ provided that A is occupied and B is unoccupied 
(denoted as $s_{k}^{(n)}(p)$, we call such clusters singly-occupied).
For convenience, A and B are not included in counting the cluster size $k$ for $t_k^{(n)}(p)$ and $s_{k}^{(n)}(p)$, but C is included.
We now introduce the generating functions $T_n(x)$ and $S_n(x)$ for $t_k^{(n)}(p)$ and $s_{k}^{(n)}(p)$ 
(Fig.~\ref{fig-diagramA}), which are defined as
\begin{subequations}
\begin{eqnarray}
T_n(x) &\equiv& \sum_{k=0}^{\infty}t_k^{(n)}(p)x^k, \\
S_n(x) &\equiv& \sum_{k=0}^{\infty}s_{k}^{(n)}(p)x^k.
\end{eqnarray}
\label{eq-gf}%
\end{subequations}
Here $T_n(1)=S_n(1)=1$ for all $n$. 
Given the self-similar structure, the recursion relations for these generating functions are readily obtained as 
\begin{subequations}
\begin{eqnarray}
T_{n+1}(x)&=& p x T_n^3(x)+q T_n(x) S_n^2(x), \label{rec-Tn} \\
S_{n+1}(x)&=& p x T_n(x)S_n^2(x)+q S_n^2(x), \label{rec-Sn}
\end{eqnarray}
\label{eq-rec}%
\end{subequations}
where $q \equiv 1-p$ (Figs.~\ref{fig-diagramB}(a) and \ref{fig-diagramB}(b)). 
The initial condition is $T_1(x)=S_1(x)=q+px$.
For example, the first term of the r.h.s. in Eq.~(\ref{rec-Tn}) represents
the contribution of the first graph of Fig.~\ref{fig-diagramB}(a) in which root C is occupied. 
It consists of the factor $p x$ for the occupied root C and the factor $T_n^3(x)$ for three doubly-occupied root clusters.
Here, $x$ accounts for the occupied root C, which is not counted in $T_n(x)$ or $S_n(x)$, 
but is counted in $T_{n+1}(x)$ or $S_{n+1}(x)$. 
The second term of the r.h.s. in Eq.~(\ref{rec-Tn}) represents
the contribution of the second graph of Fig.~\ref{fig-diagramB}(a).
The factor $q$ means the probability of root C being unoccupied, and $T_n(x) S_n^2(x)$ is given from one doubly-occupied root cluster and two singly-occupied root clusters.

Now, we consider the mean size of the root cluster. 
By $f_k^{(n)}(p)$, we denote the probability that the size of the root cluster in $G_n$ is $k$.
For evaluating the generating function  $F_n(x) \equiv \sum_{k=0}^{\infty}f_k^{(n)}(p)x^k$, 
we only need to count possible contributing diagrams as shown in Fig. \ref{fig-diagramB}(c). 
Noting that all roots A, B, and C in $G_{n+1}$ are not counted as occupied in $T_n(x)$ and $S_n(x)$, we easily find that $F_{n+1}(x)$ is evaluated as 
\begin{eqnarray}
F_{n+1}(x)&=&p^3 x^3 T_{n}^3(x)+3 p^2 q x^2 T_{n}(x)S_{n}^2(x) 
+ 3p q^2 x S_{n}^2(x) + q^3. 
\end{eqnarray}
For $n=1$, we have $F_1(x)=p^3 x^3+3 p^2 q x^2+3p q^2x+q^3$.
Then, $s_{\rm root}(N_{n+1};p)=F_{n+1}'(1)$ is given by
\begin{eqnarray}
s_{\rm root}(N_{n+1};p) &=& 3p^2T_{n}'(1) +6pq S_{n}'(1)+3p, \label{eq-recsmax}
\end{eqnarray}
and $s_{\rm root}(N_{1};p)=3p$. 
Here, the prime denotes the first derivative with respect to $x$. 
By Eq.~(\ref{eq-rec}), we have $T_{n}'(1)$ and $S_{n}'(1)$ recursively as 
\begin{subequations}
\begin{eqnarray}
T_{n+1}'(1) &=&(2p+1)T_n'(1)+2q S_n'(1)+p, \\
S_{n+1}'(1) &=& p T_n'(1)+2 S_n'(1)+p,
\end{eqnarray}
\label{eq-recursion}
\end{subequations}
with $T_1'(1)=S_1'(1)=p$. 

Now, we consider the fractal exponent of the root cluster $\psi_{\rm root}$, where $s_{\rm root}(N_n;p) \propto N_n^{\psi_{\rm root}}$. 
For $n \gg 1$, the recursion relations (\ref{eq-recursion}) are approximated as 
${\bf x_{n+1}}={\bf A}\cdot{\bf x_n}$, where 
\begin{equation}
{\bf x_n} =
\begin{pmatrix}
T_n'(1) \\
S_n'(1)
\end{pmatrix}
, \quad
{\bf A} =
\begin{pmatrix}
2p+1 && 2q \\
p && 2 
\end{pmatrix}
. \label{matrix}
\end{equation}
Solving the characteristic equation of ${\bf A}$ yields the largest eigenvalue $\lambda({\bf A})$,
\begin{equation}
\lambda({\bf A})=\frac{1}{2}(2p+3+\sqrt{1+4p-4p^2}).
\end{equation}
By noting $s_{\rm root}(N_n; p) \propto \lambda({\bf A})^{n}$, we have the fractal exponent of the root cluster as 
\begin{equation}
\psi_{\rm root}(p)=\frac{\ln (\frac{1}{2}(2p+3+\sqrt{1+4p-4p^2}))}{\ln 3}, \label{res-psi}
\end{equation}
for $0 < p \le 1$. 
At $p=0$, $\psi_{\rm root}(p)=0$.

\begin{figure}
 \begin{center}
  \includegraphics[height=55mm]{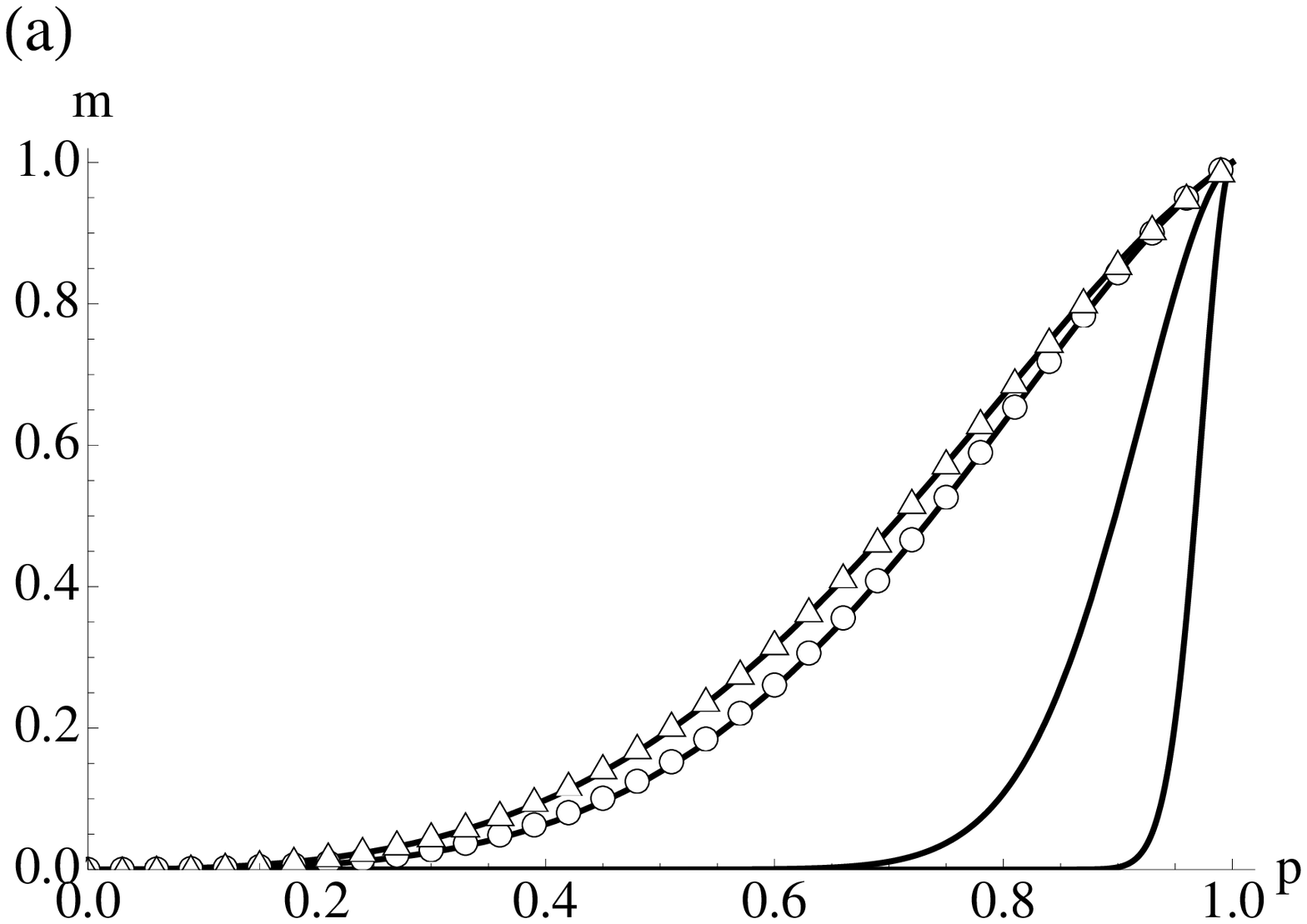}
  \includegraphics[height=55mm]{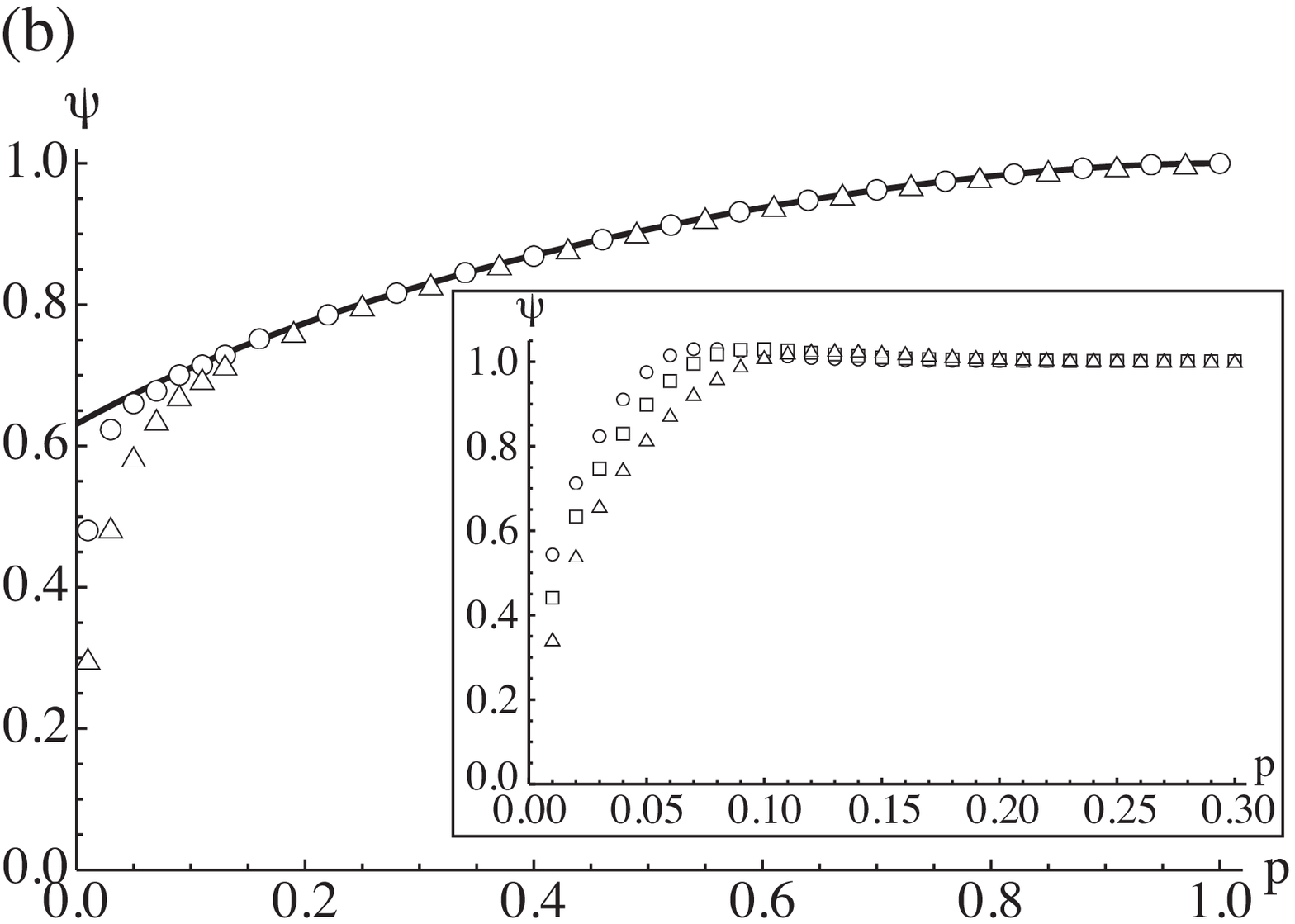}
 \end{center}
 \caption{
(a) Order parameter and (b) fractal exponent profiles for different generations.
The solid lines in (a) are the numerical evaluations of $m_{\rm root}(N_n;p)$ given 
by Eqs.~(\ref{eq-recsmax}) and (\ref{eq-recursion}) for generations $n=9$, 12, 100, and 1000 (from left to right).
The solid line in (b) represents $\psi_{\rm root}(p)$ given by Eq.(\ref{res-psi}).
The open circles and triangles denote the results obtained from the Monte Carlo simulation for (a) $m(N_n;p)$ and (b) $\psi(N_n;p)$ with $n=12$ and 9, respectively.
The inset of (b) shows the plot of the fractal exponent $\psi(N_n; p)$ of the configuration model having the same degree distribution as that of the DGM network. 
The number of nodes is $N_n=265722$ (circles), 88575 (squares), and 29526 (triangles). 
The number of trials over each realization is 1000, and the number of graph realizations is 100.
}
\label{fig-result}
\end{figure}


\section{Result}

From Eq.~(\ref{res-psi}), it is apparent that $\ln 2/\ln 3<\psi_{\rm root}<1$ for $0<p<1$.
This indicates that $p_{c1}=0$ and $p_{c2}=1$ and the order parameter is zero except at $p=1$.
In Fig.~\ref{fig-result}(a), $m_{\rm root}(N_n;p)=s_{\rm root}(N_n;p)/N_n$ is plotted over several generations using Eqs.~(\ref{eq-recsmax}) and (\ref{eq-recursion}).
We also performed the Monte Carlo simulation of site percolation
in the DGM network for different generations ($n$ ranging from 8 to 13).
The number of percolation trials at each $p$ is 200000.
The order parameters $m(N_n;p)=s_{\rm max}(N_n;p)/N_n$ obtained 
by Monte Carlo simulation for $n=9$ and 12 are shown in Fig.\ref{fig-result}(a). 
The numerical results of $m(N_n;p)$ lie precisely on the analytical curves of $m_{\rm root}(N_n; p)$.
From Fig.~\ref{fig-result}(a), we find that 
across the range of $p$, the order parameter decays to zero as $n$ increases, although the convergence is very slow.
The giant component of $O(N)$ disappears at $p<1$ in the thermodynamic limit, 
implying that this network is essentially fragile against random failures. 

In Fig.~\ref{fig-result}(b), we plot the fractal exponent $\psi_{\rm root}(p)$ given by Eq.~(\ref{res-psi}) (solid line) 
and $\psi(N_n; p)$ obtained from the Monte-Carlo simulations (symbols). 
Here, the fractal exponent $\psi(N; p)= d \ln s_{\rm max}(N;p)/d \ln N$ of a finite graph is evaluated 
as the difference $\psi(N_n; p)  \approx (\ln s_{\rm max}(N_{n+1}; p) - \ln s_{\rm max}(N_{n-1}; p))/(\ln N_{n+1}- \ln N_{n-1})$.
When $p \gtrsim 0.2$, the Monte Carlo results lie in the theoretical curve. 
When $p \lesssim 0.2$, the data points deviate from this curve, but this deviation can be diminished by increasing $n$.

The DGM network is also fragile against intentional attacks.
Note that, in $G_n$, the degree of the node added at generation $\ell$ is $2^{n-\ell}$, 
i.e., the older the node is, the larger the degree is.
Let us consider removing the nodes added at generations less than $\ell$ from $G_n$. 
Then, the three clusters including the nodes added at generation $\ell$ can be considered as the largest clusters.
Simple reasoning gives the size of these clusters as $3^{n-\ell}$ in 
$G_n$, i.e., $\psi(N_n; p)=1-\ell/n$.
To prevent the fraction of removed nodes $\tilde{p}=1-p=3^{\ell-1}/3^n$ from disappearing in the limit $n \to \infty$,
$\ell$ needs to increase as $\ell=n-c$, where $c$ is some constant.
Then we conclude that $\psi=c/n \to 0$ for $p<1$, implying that $p_{c1}=1$. 
Thus, we have $p_{c1}=p_{c2}=1$ for intentional attacks.


\section{Summary}

To summarize, we have examined site percolation in the DGM network. 
We have shown that $p_{c1}=0$ and $p_{c2}=1$ for site percolation (random failure), 
while $p_{c1}=p_{c2}=0$ for bond percolation \cite{dorogovtsev2003renormalization}.
We also have demonstrated that $p_{c1}=p_{c2}=1$ for intentional attacks.
We conclude that this hierarchical network is fragile against both random failure and intentional attacks. 

How universal are the behaviors observed here among complex networks?
The origin of the observed fragility should be assigned to the hierarchical structure of the DGM network.
We have numerically observed site percolation in the configuration model having the same degree distribution as that of the DGM network (see the inset of Fig.~\ref{fig-result}(b)). 
The numerically obtained $\psi(N; p)$ approaches 1 even for small $p(>0)$ when $N$ increases. 
This means $p_{c2}=0$, which is consistent with the result by the local tree approximation \cite{callaway2000robustness,cohen2000resilience} (note that degree exponent $\gamma$ of this network is less than 3).
Furthermore, for site-bond percolation in another hierarchical scale-free network, called the decorated (2,2) flower, 
the critical open bond probability is 1 except for the case of no dilution \cite{hasegawa2012phase}. 
An important point here is that the difference in fragility between bond and site percolations in the DGM network 
may be related to a geometrical property, i.e., the number of {\it ends}. 
Consider an infinite graph $G$. 
The number of ends of $G$, $e(G)$, is given as the supremum of the number of infinite connected components in $G \backslash S$, 
where $G \backslash S$ is the graph obtained from $G$ by removing an arbitrary finite subset $S$ of nodes or edges. 
If $G$ is locally finite and transitive, $p_c=1$ when $e(G)=\infty$
\footnote{{The number of ends of an infinite, locally finite, transitive graph is either of 1, 2, or $\infty$} \cite{mohar1991some}.}  
and $p_c<1$ when $e(G)=1$
\footnote{{There is a widely-believed conjecture that $p_c<1$ if $e(G)$ of an infinite, locally finite, transitive graph $G$ is one} \cite{lyons2000phase}.}.
In the infinite DGM network $G_\infty$, which is neither locally finite nor transitive, 
the number of ends for the deletion of {\it nodes} and that for the deletion of {\it edges} can be considerably different:
In the case of node and edge deletions, $e(G_\infty)$ is infinite and unity, respectively.
In other words, $G_\infty$ can disintegrate 
if a finite number of nodes is removed, which occurs with a nonzero probability when $p<1$, 
but is robust against edge removals, thus reflecting the qualitative difference in percolation threshold between site and bond percolations.
Such a structural defect appears to be common among previously identified hierarchical scale-free networks, 
such as the (decorated) ($u,v$) flower, the Ravasz-Barab{\'a}si hierarchical network \cite{ravasz2003hierarchical}, 
and a hierarchical network proposed by Barab{\'a}si {\it et al.} \cite{barabasi2001deterministic}.
We expect that fragility against random failures is also common in hierarchical network models.
The validity of this conjecture will be investigated in future studies.


\section*{Acknowledgements}
TH thanks T.\ Kondo, R.\ Tanaka, and T.\ Nogawa for fruitful discussions.
TH acknowledges the support through Grant-in-Aid for Young Scientists (B) (Grant No.\ 24740054) from MEXT, Japan.
This research was partially supported by JST, ERATO, Kawarabayashi Large Graph Project. 


\end{document}